\documentclass[fleqn,usenatbib,manuscript]{aastex631}

\usepackage{threeparttable}
\usepackage{textcomp,booktabs}
\usepackage{multirow}
\usepackage{enumitem}

\usepackage{newtxtext,newtxmath}
\usepackage[T1]{fontenc}

\usepackage{natbib}
\DeclareRobustCommand{\VAN}[3]{#2}
\let\VANthebibliography\thebibliography
\def\thebibliography{\DeclareRobustCommand{\VAN}[3]{##3}\VANthebibliography}

\usepackage{graphicx}	
\usepackage{amsmath}

\usepackage{amssymb}
\usepackage{gensymb}
\usepackage{color}
\usepackage{mathtools}
\usepackage{verbatim}

\begin{document}
\title{An atypical low-frequency QPO detected in the hard state of MAXI J1348$-$630 with $Insight$-HXMT}
\correspondingauthor{Zhen Yan \& Ren-Yi Ma}
\email{zyan@shao.ac.cn, ryma@xmu.edu.cn}
\author{Xin-Lei Wang}
\affil{Department of Astronomy and Institute of Theoretical Physics and Astrophysics, Xiamen University, Xiamen, Fujian 361005, China}

\author{Zhen Yan}
\affil{Key Laboratory for Research in Galaxies and Cosmology, Shanghai Astronomical Observatory, Chinese Academy of Sciences, 80 Nandan Road, Shanghai 200030, China}
\affil{SHAO-XMU Joint Center for Astrophysics,  Xiamen, Fujian 361005, China}
\author{Fu-Guo Xie}
\affil{Key Laboratory for Research in Galaxies and Cosmology, Shanghai Astronomical Observatory, Chinese Academy of Sciences, 80 Nandan Road, Shanghai 200030, China}
\affil{SHAO-XMU Joint Center for Astrophysics,  Xiamen, Fujian 361005, China}
\author{Jun-Feng Wang}
\affil{Department of Astronomy and Institute of Theoretical Physics and Astrophysics, Xiamen University, Xiamen, Fujian 361005, China}
\affil{SHAO-XMU Joint Center for Astrophysics,  Xiamen, Fujian 361005, China}
\author{Ren-Yi Ma}
\affil{Department of Astronomy and Institute of Theoretical Physics and Astrophysics, Xiamen University, Xiamen, Fujian 361005, China}
\affil{SHAO-XMU Joint Center for Astrophysics,  Xiamen, Fujian 361005, China}

\begin{abstract}
Based on the \textit{Insight}-HXMT archival data, we have detected a new atypical low-frequency quasi-periodic oscillation (LFQPO) in the black hole X-ray binary MAXI J1348$-$630. The new LFQPO is detected in all the three instruments of $Insight$-HXMT with a combined significance of 3--5 $\sigma$, covering a wide energy range of 1--100 keV. The fractional root-mean-square (RMS) seems decrease with energy. It exclusively appears in the hard state during both the main and mini outburst, spanning an X-ray intensity range by a factor of 10, and a very narrow hardness range. The frequency of this new type of LFQPO is moderately stable, in the range of 0.08--0.15 Hz. We discussed different models for the LFQPO, and found none is able to explain the observed properties of this new type of LFQPO.

\end{abstract}



\section{Introduction}
\label{sec:intro}

As we know, the luminosity and X-ray spectrum of the black hole X-ray binaries (BHXRBs) usually go through different stages during the outbursts, i.e., from the hard state (HS), to hard-intermediate state (HIMS) and soft-intermediate states (SIMS), then to the soft state (SS) during the outburst rise, and then back to the hard state during the outburst decay (for reviews, see e.g., \citealt{Homan_2005,Remillard_2006,Belloni_2016}).
During the outburst, the variation of the X-ray intensity and hardness generally follows a q-shape trajectory in the hardness-intensity diagram (HID, e.g. \citealt{Homan01}).


In addition to the above long-term variation, BHXRBs also exhibit short-term variations all along the outburst, which can be studied by the power density spectrum (PDS). The PDS is typically dominated by the broad band noise, but sharp peaks sometimes emerge, which correspond to quasi-periodic oscillations (QPOs) and reveal characteristic time scales of the accretion flow around the black hole\citep[e.g.,][]{Nowak00, Casella04, Belloni_2014,Ingram20}.
Traditionally, QPOs have been quantitatively described using three parameters, i.e., the central frequency, the quality factor Q (defined as the ratio of the center frequency to the full width at half maximum, FWHM), and the fractional RMS.
According to the central frequency, QPOs are classified into two categories: low-frequency QPOs (LFQPOs),  which is in the range of a few mHz up to $\sim$10 Hz \citep[e.g., ][]{Belloni02a, Casella04}, and high-frequency QPOs (HFQPOs), which is in the range of tens to the hundreds of Hz \citep[e.g., ][]{Morgan97, Homan01,Remillard_2006,Belloni_2014}.

The LFQPOs have been widely studied. They are further classified into three types, i.e. types-A, B, and C, based on the parameters, the corresponding X-ray spectral state, and the characteristics of the broad band noise \citep{Casella04, Casella05, Motta_2011}. The type-C QPO is the most common LFQPO detected in the BHXRBs. 
It usually is the prominent feature in the PDS during HS and HIMS, being accompanied by a band-limited noise \citep[BLN; e.g.][]{Belloni02a}. The frequency of type-C QPO usually increases from tens of mHz to $\sim$10 Hz \citep[e.g., ][]{Motta_2011,Buisson2019}. In addition, type-C QPO is relatively narrow, and the Q value is high. The mechanism underlying the formation of type-C QPOs is not fully understood yet. The present physical models can be divided into two categories: the instability of the accretion flow \citep[e.g., ][]{Tagger1999, Cabanac10} and the geometrical effects\citep[e.g., ][]{Stella98, Stella99,Ingram09, Ingram11,Veledina_2013}. Recent studies on fractional RMS and phase lag of the type-C QPO have shown that the geometric origin is more promising \citep[e.g., ][]{Motta_2015,vandenEijnden2017,You2018}.

The type-B QPO is detected in the SIMS, with the central frequency at 1--7 Hz. It is characterised by the relatively strong 4--5\% fractional RMS amplitude and the narrow peak \citep[e.g, ][]{Wijnands99, Casella04, Casella05,Belloni_2014}. 
The background noise is weak power-law or broadband. Sometimes, the weak harmonics can be detected.
Similarly, the type-A QPO also appears in the SIMS state as a weak and broad peak. But it is the least common LFQPO \citep{Wijnands99,Homan01,Casella04,Belloni_2014}.


MAXI J1348$-$630 was discovered in its 2019 outburst \citep{2019ATel12425....1Y} and its location was subsequently determined \citep{2019ATel12434....1K}. Based on the spectral-timing analysis of the observational data by the Neutron Star Interior Composition Explorer (\textit{NICER}),
the source is identified as a black hole system \citep{2020MNRAS.499..851Z}.
The three types of LFQPOs have been reported during the main and mini outbursts: type-A QPO with the frequency of $\sim7$ Hz \citep{2020MNRAS.499..851Z,zhang_2023}, type-B QPO with the frequency about 4--5 Hz \citep{2020MNRAS.496.4366B,2020MNRAS.499..851Z,2021MNRAS.505.3823Z}, and type-C QPO with the frequency in the range from 0.2 Hz to 7 Hz \citep{2020MNRAS.499..851Z,2021MNRAS.505..713J}. 
The variability and phase lags of the type-C QPO during the main and mini outburst have been studied by \citet{Alabarta_2022}.
Moreover, a HFQPO at $\sim$98 Hz is detected during the mini outburst \citep{Kumar2023}.

In this paper, we report the detection of a new type of LFQPO based on the observations by \textit{Insight}-HXMT observations on MAXI J1348$-$630. In the following, we present the observation and data reduction in \autoref{sec:data}, the data analysis and results in \autoref{sec:analysis results}, and discuss the properties and possible mechanism of the new type of LFQPO in \autoref{sec:discussion}.


\section{OBSERVATION AND DATA REDUCTION}
\label{sec:data}

The \textit{Insight}-HXMT is an X-ray astronomical satellite with the detectable energy range of 1--250 keV. There are three payloads: the Low Energy payload (LE, 1--12 keV), the Medium Energy payload (ME, 8--35 keV) and the High Energy payload (HE, 20--250 keV), with the time resolution of 1 ms, 240 $\mu$s, 4 $\mu$s and the effective area of 384 cm$^{2}$ , 952 cm$^{2}$ , 5100 cm$^{2}$, respectively \citep{zhang_overview_2020}.
The data segment from \textit{Insight}-HXMT is named as ``exposure'' and identified as Exposure ID. The duration of each exposure lasts approximately three hours.

The \textit{Insight}-HXMT performed a monitoring campaign on MAXI J1348$-$630 from 2019 January 27 to 2019 July 29, which covers a main and a mini outburst \citep{Alabarta_2022}.
We have analyzed a total of 288 \textit{Insight}-HXMT exposures in this work.

The data were processed by using the \textit{Insight}-HXMT  Data Analysis Software package (\texttt{HXMTDAS}) V2.05.
After calibrating and screening, the raw event data are processed to screened event files.
Then we used light curves extraction tools (\texttt{helcgen,melcgen,lelcgen} in \texttt{HXMTDAS}) to generate light curves with a time resolution of 1/256 seconds  for LE, ME, HE instruments, respectively.
The background light curves with the same resolution and the same energy band are produced by using \texttt{lebkgmap}, \texttt{mebkgmap} and  \texttt{hebkgmap}, respectively. The net light curves are then calculated by using \texttt{lcmath}.
The averaged PDS with a frequency resolution of 1/128 Hz can be generated with \texttt{powspec} by using the net light curves.
Due to short good time intervals (GTIs) in some exposures, we also used a 64s duration with a 0.0156 Hz resolution to obtain more segments to average the PDS.


\begin{figure*}
\includegraphics[width=1\linewidth]{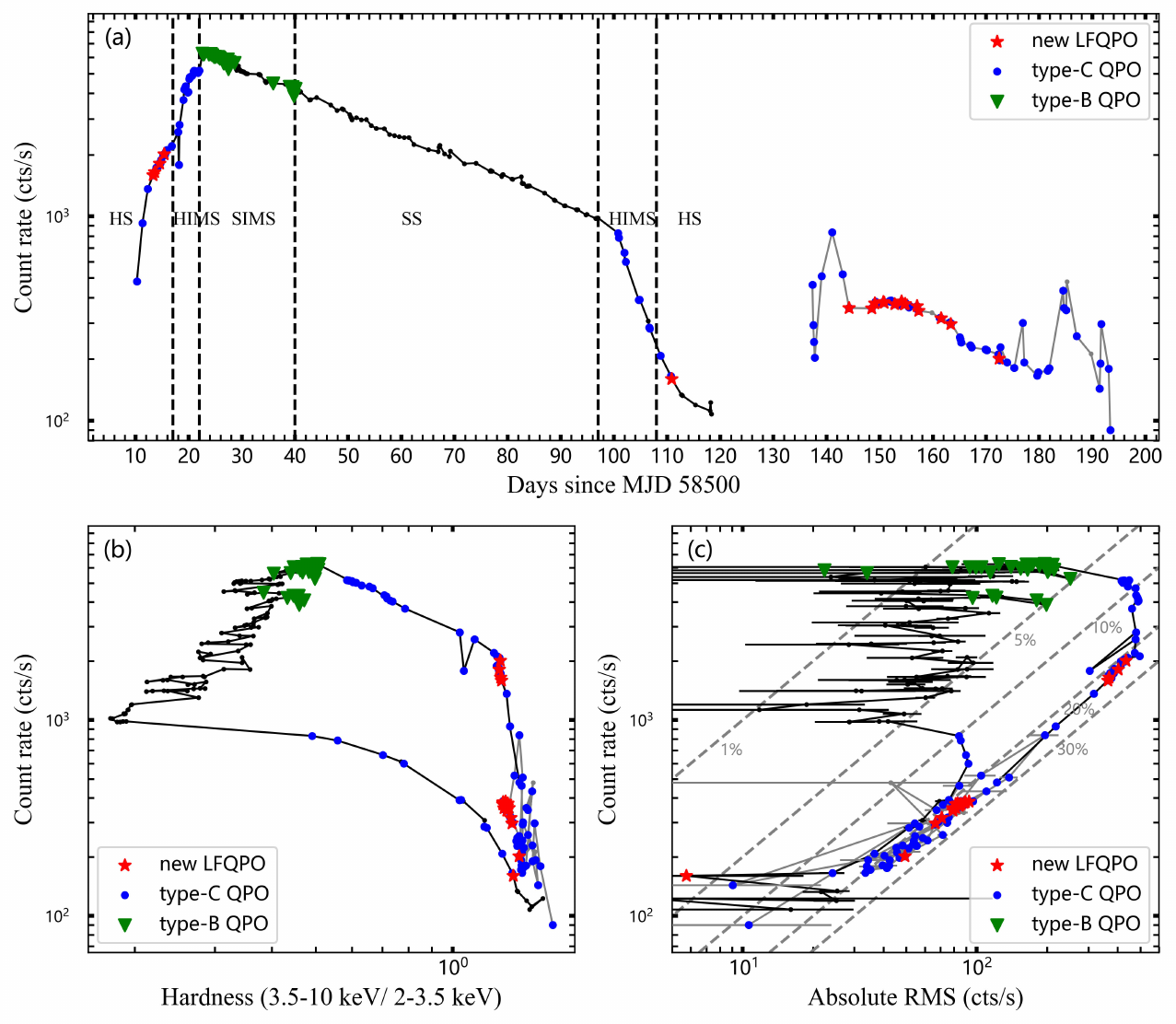}
\caption{(a)The light curve of MAXI J1348$-$630, using the photon count rate from the LE of \textit{Insight}-HXMT in the 1--10 keV energy range. The time reference point is set at MJD 58500. The  black line and the grey lines represent the main and mini outbursts, respectively. (b) The hardness-intensity diagram of MAXI J1348$-$630. The count rate is from the 1--10 keV energy band, and the hardness is calculated as the ratio of photon counts in the 3.5--10 keV band over those in the 2--3.5 keV band. (c)The RMS-intensity diagram, wherein the RMS was calculated over a frequency range of 0.5--64Hz, with the count rate obtained from the 1–10 keV energy band. In all diagrams, red stars, blue circles, and green triangles denote the new LFQPOs, type-C QPOs, and type-B QPOs, respectively.}
\label{fig:lc-hid-rid}
\end{figure*}

\section{DATA ANALYSIS AND RESULTS}
\label{sec:analysis results}

\subsection{Model fitting of PDS}
All PDS derived from \textit{Insight}-HXMT were applied the Leahy normalization \citep{Leahy1983}, in which the Fourier transform function of the light curve is normalized by the total photon count. We then formatted the PDSs to be compatible with \texttt{XSPEC} (version 12.10.1) using the tool \texttt{flx2xsp} \citep{Ingram12}, which allows us to employ models within \texttt{XSPEC} for the data fitting purpose.

We fitted all the PDSs with the combination model of \texttt{powerlaw} and/or  \texttt{lorentz} in \texttt{XSPEC} and used chi-square statistics to assess the goodness-of-fit. The value and the 1 $\sigma$ range of the best-fitting parameters are retrieved from the chains generated from the Monte Carlo Markov Chain (MCMC) algorithm \texttt{emcee} implemented in \texttt{XSPEC} \footnote{\url{https://github.com/zoghbi-a/xspec_emcee}}. In most exposures, three \texttt{lorentz} components, i.e., two zero-centered and one broad, are used to fit the PDSs. Sometimes one or several narrow \texttt{lorentz} components are required. The zero-centered and narrow \texttt{lorentz} components are used to describe the BLN and QPO, respectively.
For clarity, we refer to the lower-frequency BLN as BLN1 and the other as BLN2.
The spectral index of one \texttt{powerlaw} was fixed at 0 to account the white noise. The best-fitting normalization of this \texttt{powerlaw} component is always around 2 (1.97 -- 2.03), which is consistent with the white noise in Leahy normalization.

We used the hardness-intensity diagram (HID) and the RMS-intensity diagram (RID) to demonstrate the spectral and timing evolution of MAXI J1348$-$630 in both main and mini outbursts. Using the LE of \textit{Insight}-HXMT, we calculated the hardness between 3.5--10 keV and 2--3.5 keV energy bands, and count rate (1--10 keV) to plot the HID as demonstrated in \autoref{fig:lc-hid-rid}b. As for the RID, we also used data from LE in 1--10 keV energy band. We first integrated the power within the 0.5--64 Hz frequency range of the Leahy normalized PDS after subtracting white noise. We then calculated the absolute RMS by using the integrated power and the mean count rate as  $\sqrt{P_{0.5-64Hz}\times R_{mean}}$.

As shown in \autoref{fig:lc-hid-rid}b, the main outburst of MAXI J1348$-$630 displays the typical  ``q'' shape of a black hole X-ray binary \citep{munoz2011,Belloni_2016}. The different states are clearly determined from the observed changes in intensity, hardness, and absolute RMS. Before MJD 58517, the outburst was in the hard state, during which the absolute RMS increases linearly with count rate. Significant type-C QPOs with increasing central frequencies were observed. 
The source subsequently transitioned into the hard-intermediate state, characterized by a rapid rise in intensity and a corresponding decrease in both hardness and RMS. 
After MJD 58522, the hardness and RMS continued to decrease, and the type-B QPO appeared, signifying the transition into the soft-intermediate state.
After MJD 58540, the source converted to the soft state, and persisted until MJD 58597. Then the hardness and RMS increased with decreasing count rate, indicating a transition from the soft state back to the hard state. The detailed spectral and timing evolution during this stage have been reported in \citet{2020MNRAS.499..851Z} and \citet{Alabarta_2022} based on the $NICER$ data. The mini outburst follows the track of the hard state in the HID and RID \citep[see also ][]{Carotenuto2021MNRAS}.


\subsection{The detection of a new LFQPO at $ \sim 0.1 $ Hz }
\label{sec:PDS examples}

\begin{figure*}
\resizebox{1\columnwidth}{!}{\rotatebox{0}{\includegraphics[clip]{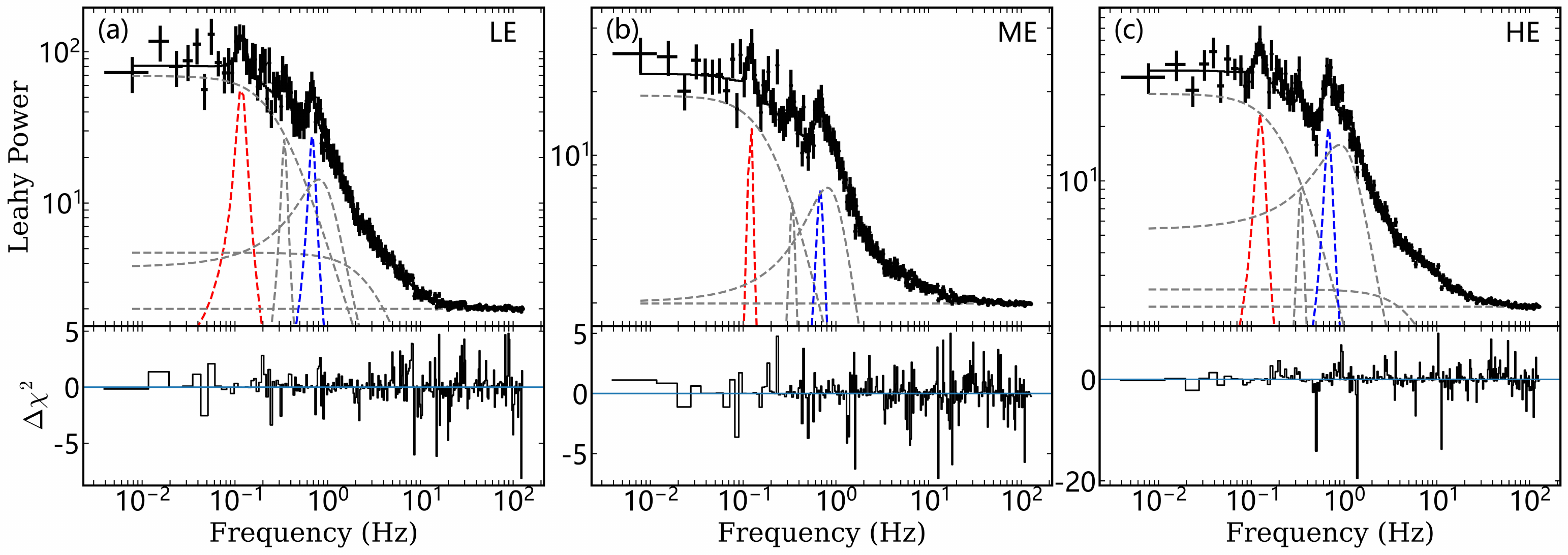}}}
\caption{The power spectra obtained from three instruments of the $Insight$-HXMT for expID P021400200401 at MJD 58513.30. We applied the same model to fit the three power spectra, and the fitting results are listed in \autoref{tab:pds-par}. The blue dashed lines represent the type-C QPOs. A new LFQPO with a frequency around 0.12 Hz is present in all power spectra, indicated by the red dashed lines. The BLNs and the sub-harmonic of the type-C QPOs are represented by the grey dashed lines.}
\label{fig:example-pds}
\end{figure*}

\begin{table*}
  \centering
  \caption{The fitting parameters in \autoref{fig:example-pds}. The absence of the error for the sub-harmonic frequency is because we fixed it at half the value of the type-C QPO.}
 \begin{tabular}{rllll}
    \toprule
    \multicolumn{2}{c}{} & \multicolumn{1}{c}{HXMT LE} & \multicolumn{1}{c}{HXMT ME} & \multicolumn{1}{c}{HXMT HE} \\
    \midrule
    \multicolumn{1}{c}{new LFQPO} & Frequency (Hz) & $0.118^{+0.005}_{-0.005}$ & $0.121^{+0.003}_{-0.003}$ & $0.125^{+0.005}_{-0.004}$ \\
          & FWHM (Hz) & $0.030^{+0.020}_{-0.012}$ & $0.014^{+0.010}_{-0.006}$ & $0.024^{+0.017}_{-0.010}$ \\
          & RMS (\%) & $4.43^{+0.84}_{-0.84}$ & $1.93^{+0.44}_{-0.34}$ & $1.55^{+0.30}_{-0.26}$ \\
    \midrule
    \multicolumn{1}{c}{type-C QPO} & Frequency (Hz) & $0.684^{+0.009}_{-0.010}$ & $0.679^{+0.012}_{-0.012}$ & $0.679^{+0.008}_{-0.009}$ \\
          & FWHM (Hz) & $0.100^{+0.046}_{-0.032}$ & $0.153^{+0.060}_{-0.050}$ & $0.124^{+0.031}_{-0.029}$ \\
          & RMS (\%) & $5.23^{+0.99}_{-0.90}$ & $4.11^{+0.83}_{-0.78}$ & $3.14^{+0.43}_{-0.34}$ \\
    \midrule
    \multicolumn{1}{c}{sub-harmonic} & Frequency (Hz) & $0.342^{+0.0}_{-0.0}$ & $0.340^{+0.0}_{-0.0}$ & $0.339^{+0.0}_{-0.0}$ \\
    \multicolumn{1}{c}{of type-C QPO} & FWHM (Hz) & $0.052^{+0.031}_{-0.023}$ & $0.065^{+0.031}_{-0.026}$ & $0.051^{+0.036}_{-0.024}$ \\
          & RMS (\%) & $3.57^{+0.88}_{-0.72}$ & $2.47^{+0.54}_{-0.45}$ & $1.35^{+0.37}_{-0.33}$ \\
    \midrule
    \multicolumn{1}{c}{BLN} & Frequency (Hz) & $0.288^{+0.070}_{-0.063}$ & $0.214^{+0.036}_{-0.029}$ & $0.222^{+0.038}_{-0.032}$ \\
          & RMS (\%) & $13.3^{+1.5}_{-1.6}$ & $8.50^{+0.63}_{-0.59}$ & $5.37^{+0.45}_{-0.38}$ \\
    \midrule
    \multicolumn{1}{c}{broad lorentz} & Frequency (Hz) & $0.77^{+0.13}_{-0.10}$ & $0.826^{+0.063}_{-0.061}$ & $0.903^{+0.061}_{-0.048}$ \\
          & FWHM (Hz) & $0.96^{+0.18}_{-0.15}$ & $1.028^{+0.091}_{-0.083}$ & $1.272^{+0.075}_{-0.071}$ \\
          & RMS (\%) & $11.1^{+2.1}_{-2.2}$ & $9.94^{+0.74}_{-0.73}$ & $8.41^{+0.35}_{-0.41}$ \\
    \midrule
    \multicolumn{1}{c}{BLN} & Frequency (Hz) & $3.40^{+0.38}_{-0.29}$ & $7.49^{+0.55}_{-0.59}$ & $7.75^{+0.33}_{-0.30}$ \\
          & RMS (\%) & $12.58^{+0.46}_{-0.52}$ & $10.84^{+0.22}_{-0.18}$ & $9.07^{+0.10}_{-0.10}$ \\
    \midrule
     & $\chi^{2}/dof$ & 280.7/277 &271.1/277 &364.8/277 \\
    \bottomrule
    \bottomrule
    \end{tabular}%
  \label{tab:pds-par}%
\end{table*}%

When we checked the PDSs of all exposures, we noticed a peak around 0.1 Hz in the PDSs for some exposures, which have not been reported in previous works and should be a newly detected LFQPO. The representative PDSs with the new LFQPO are shown in \autoref{fig:example-pds}, which are derived from  expID P021400200401 at MJD 58513.30 according to all the three instruments of \textit{Insight}-HXMT. 
The same model as mentioned in previous subsection is applied to fit all the three PDSs. Two zero-centered Lorentzians are utilized to fit the BLNs, and three narrow ones are for the QPOs. Two of the three QPOs are type-C at $\sim 0.68$ Hz and its sub-harmonic, and the third is the new LFQPO at $\sim 0.12$ Hz (see \autoref{fig:example-pds} and \autoref{tab:pds-par}).  There is an additional component around 1 Hz, which is accounted by the broad Lorentzian component.
All the best-fitting results are listed in \autoref{tab:pds-par}.

An empirical quantity for evaluating the significance of the QPO is the signal-to-noise ratio of the integral of its power, which is represented by the normalization of the Lorentzian model \citep[e.g.][]{Motta_2015}. 
So the significance of the QPO is given as the best-fitting parameter \texttt{norm} dividing by its 1$\sigma$ uncertainty. The RMS is calculated as square root of the ratio between \texttt{norm} and the mean count rate. So the signal-to-noise ratio of RMS is twice of the \texttt{norm}.
From the fitting results in \autoref{tab:pds-par}, the new LFQPO is detected at a significance of 2.64 $\sigma$, 2.84 $\sigma$ and 2.98 $\sigma$ for LE, ME and HE, respectively, which gives a combined detection significance of 4.57 $\sigma$ by using the Fisher method \citep{Fisher1925}.

In total, the new LFQPO has been significantly detected in twenty exposures of \textit{Insight}-HXMT (\autoref{fig:lc-hid-rid}a) with a combined significance exceeding 3 $\sigma$, exhibiting frequencies ranging from 0.08 to 0.15 Hz.
All the detailed information of the new LFQPOs are listed in \autoref{tab:all-fit-value}. We then further explore the properties of the new LFQPOs and demonstrate the differences to the typical LFQPOs.

To determine the spectral states of the detection of the new LFQPO, we have presented the appearances of the new LFQPO and type-C and type-B QPOs on both the HID (\autoref{fig:lc-hid-rid}b) and the RID (\autoref{fig:lc-hid-rid}c). As illustrated in both diagrams, the new LFQPOs exclusively appear in the hard state during both the main outburst and the mini outburst (\autoref{fig:lc-hid-rid}a). The observations with significant detection of new LFQPO span a very narrow range of hardness and a large range of count rate (\autoref{fig:lc-hid-rid}b).

\begin{figure*}
\resizebox{1\columnwidth}{!}{\rotatebox{0}{\includegraphics[clip]{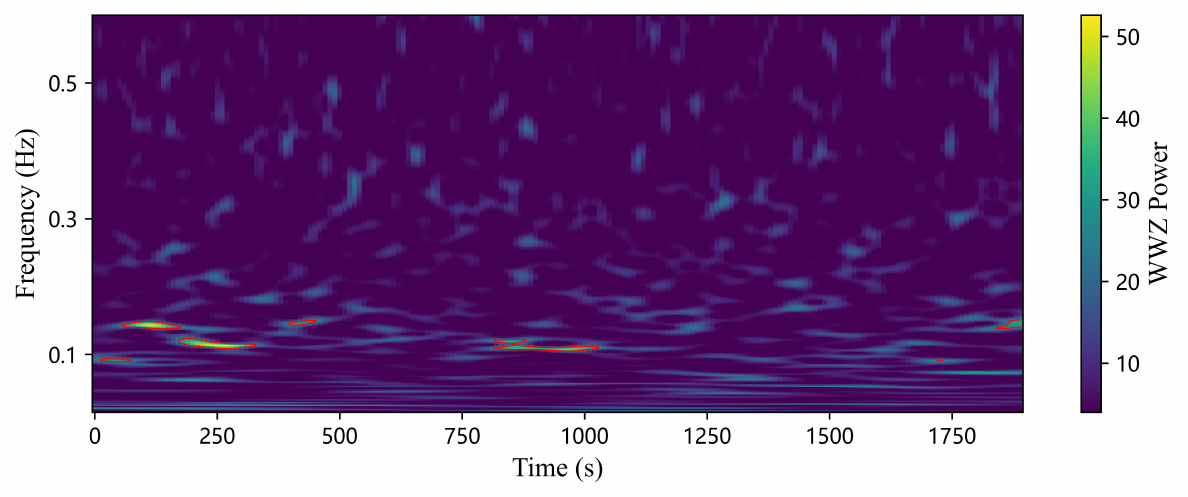}}}
\caption{Dynamic WWZ power in the time-frequency space produced with HE data during expID P020101200803. The red contours represent the durations with QPO signals $> 3 \sigma$.}
\label{fig:WWZ}
\end{figure*}

The appearance of the new LFQPO seems to be intermittent, which is common among different types of LFQPOs \citep{Middleton2011,Urquhart2022,vandenEijnden2016,Yan2021,Zhang2023}. The new LFQPO is not significant in all exposures. Except the 20 exposures with new LFQPO detection $>3\sigma$, the narrow feature around 0.1 Hz is weak or absent in other exposures. The combined significances of those exposures are all below $3\sigma$ when a narrow Lorentzian model is applied. So the detection of the new LFQPO in these exposures is insignificant. We calculated the average fractional RMS of the narrow Lorentzian model for both significant and insignificant detections. The results show that in the LE, ME, and HE bands, the average fractional RMS for significant detections are $3.92 \pm 0.98\%$, $2.06 \pm 0.45\%$, and $1.56 \pm 0.34\%$, respectively, while for insignificant detections, these values are $2.48 \pm 0.78\%$, $1.33 \pm 0.44\%$, and $1.04 \pm 0.33\%$, respectively. Notably, the former values are significantly higher than the latter, suggesting that a decrease in QPO amplitude correlates with to a reduction in detection significance.

Furthermore, the new LFQPO is intermittent within an individual exposure. To illustrate this phenomena, we employed the weighted wavelet Z-transform (WWZ) method \footnote{\url{https://github.com/eaydin/WWZ}}, which is suitable for studying the time evolution of the parameters of the signals, such as the frequency, amplitude and phase \citep{1996AJ....112.1709F,Zhang2023}. As shown in \autoref{fig:WWZ}, we present the dynamic WWZ power in the time-frequency space during expID P020101200803 with HE data. The central frequency and the FWHM of new LFQPO is 0.127 Hz and 0.038 Hz in this exposure. The number of segments was set at 100, and the frequency resolution was set at 0.002 Hz. 
Within an exposure, the significance of new LFQPO features is determined by comparing their power with the distribution of power values within the FWHM of the central frequency \citep[e.g. ][]{Zhang2023}. 
We marked all new LFQPO durations with a significance higher than $3\sigma$ (red contours of \autoref{fig:WWZ}), i.e. the power exceeds the average power more than 3 times of the standard deviation. It can be seen that the new LFQPO durations is obviously discontinuous. So we conclude that the amplitude of the new LFQPO changes not only for different exposures but also for different periods within an exposure, which demonstrates that the new LFQPO appears intermittently.

\subsection{Frequency evolution of the new LFQPO}
\label{sec:stable-frequency}

\begin{figure*}
\includegraphics[width=1\textwidth]{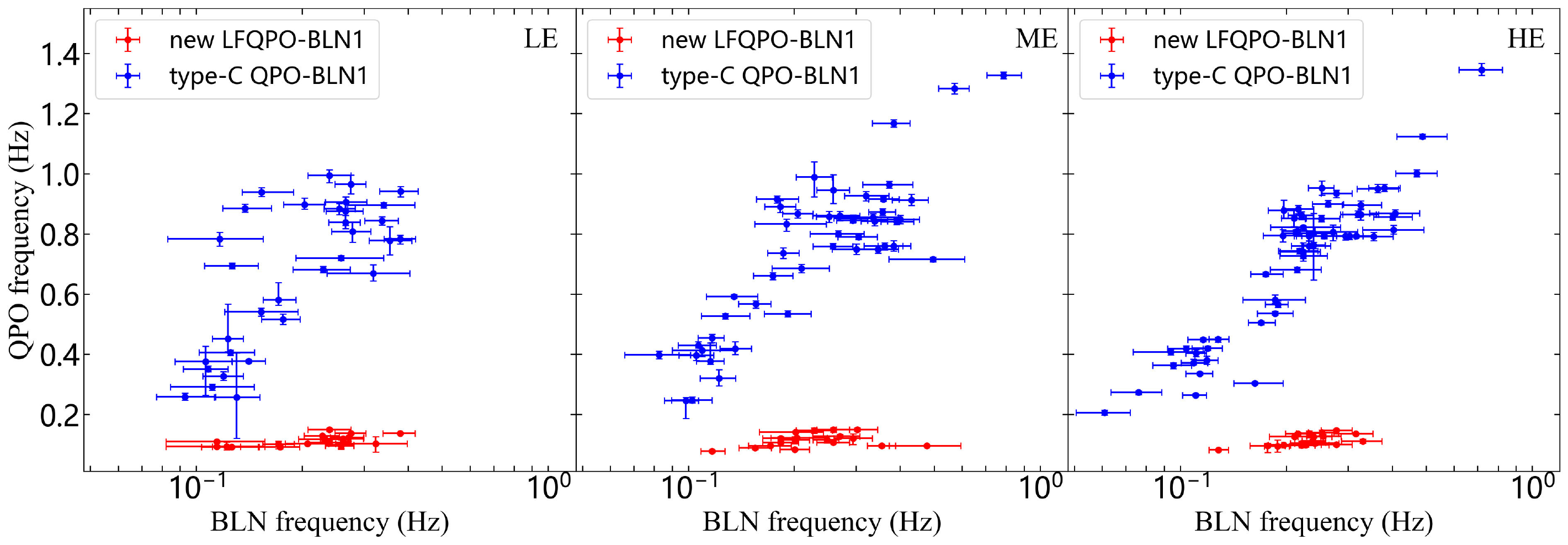}
\caption{The relationship between the frequencies of the QPOs and the lower frequency BLN. The data, from left to right, correspond to the LE, ME, and HE detectors of the \textit{Insight}-HXMT.}
\label{fig:QPOfre-BLNfre}
\end{figure*}

The new LFQPO exhibits a narrow frequency range, consistently staying between 0.08 and 0.15 Hz. As previously mentioned, this new LFQPO was only observed during the hard state. In contrast, type-C QPOs display a much broader frequency range during the hard state, spanning from 0.28 to 1.34 Hz. We then proceeded to compare the frequency evolution of these two types of QPOs.

There is a well-known correlation between the frequencies of the LFQPO and BLN among different XRBs \citep{psaltis_1999,wijnands_1999}. For QPOs, we used their central frequencies, and for BLN, we used their characteristic frequencies, which is calculated using the formula $\nu_{\text{max}} = \sqrt{{\nu_0}^2 + \Delta^2}$ \citep{Belloni_2002}, where $\nu$ is the central frequency and $\Delta$ is the half width at half maximum (HWHM). We then plotted the correlation between the frequencies of the QPOs (including the new LFQPO and the type-C QPO) and BLN1 in \autoref{fig:QPOfre-BLNfre}. Since the new LFQPO appears exclusively during the hard state, we only depicted the type-C QPO frequencies corresponding to this period. The frequencies of type-C QPO and BLN1 follow a positive correlation similar to previous works \citep{psaltis_1999,wijnands_1999}. The new LFQPO, however, obviously deviates from the correlation, and remains almost constant.

\subsection{Energy-dependence of the new LFQPO}
\label{sec:Energy-dependence}


\begin{figure*}
\includegraphics[width=1\textwidth]{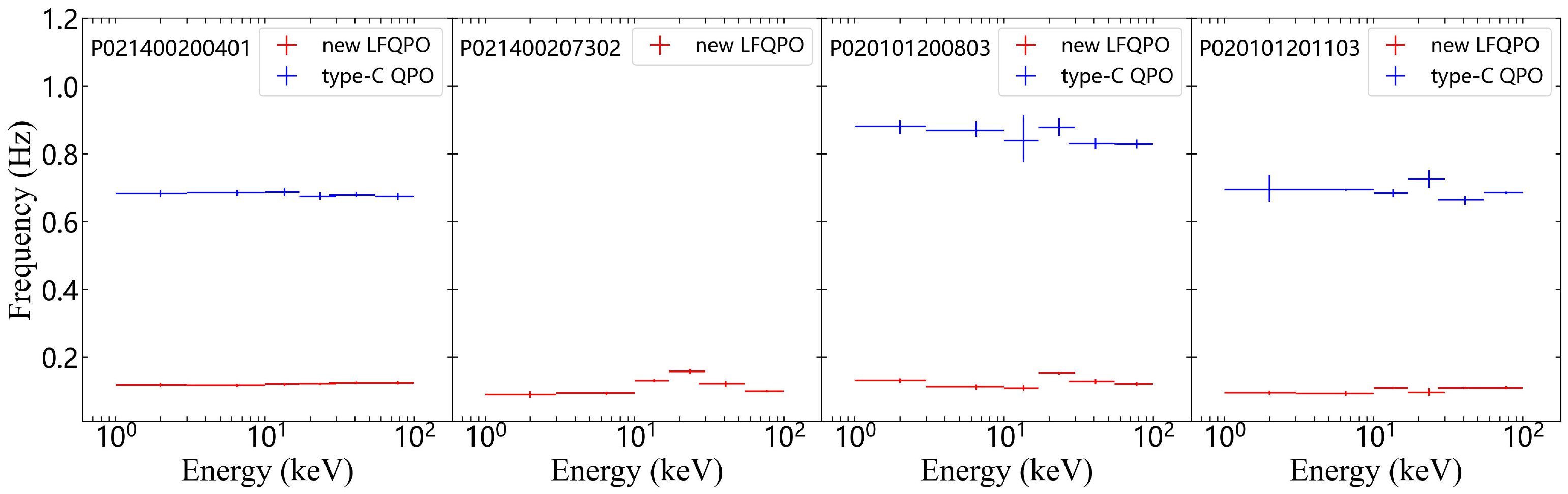}
\caption{Energy-dependence of frequencies of new LFQPOs and the type-C QPOs.}
\label{fig:fre-energy}
\end{figure*}

\begin{figure*}
\includegraphics[width=1\textwidth]{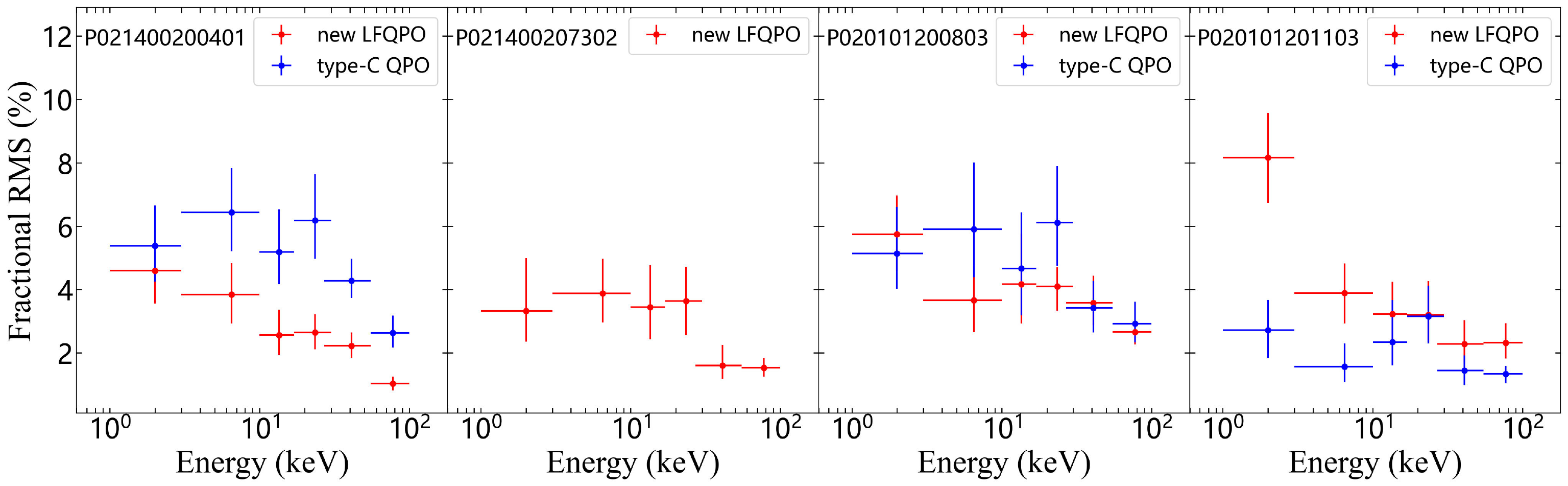}
\caption{Energy-dependence of fractional RMS of new LFQPOs and the type-C QPOs.}
\label{fig:rms-energy}
\end{figure*}

In the majority of the 20 exposures with significant new LFQPO detection, as shown in \autoref{tab:all-fit-value}, the central frequency of the new LFQPO remains relatively stable among the different energy bands of the three instruments, while the fractional RMS generally decreases from LE to HE. In order to explore the energy dependence of the frequency and fractional RMS for the new LFQPO, we generated and fitted the PDSs in different energy bands: 1--3 keV, 3--10 keV from LE, 10--17 keV, 17--30 keV from ME, and 27--55 keV, 55--100 keV from HE, respectively.
We used the four exposures with the individual significance from the LE, ME, and HE all exceeding 2 $\sigma$. We also plotted the results of type-C QPO for comparison when the type-C QPO is detected (see \autoref{fig:fre-energy} and \autoref{fig:rms-energy}).

As shown in \autoref{fig:fre-energy}, the frequency variation of the new LFQPO is usually very small across a broad energy band of 1--100 keV, which is similar to the type-C QPO, except for expID P021400207302, where the central frequency increases to 0.158 Hz at 17-30 keV and then decreases to 0.099 Hz at 55-100 keV. We also notice that other two exposures (expID P021400200410 and P020101200701) exhibit similar frequency variations, with the highest frequencies appearing at 17-30 keV. Overall, the central frequency of the new LFQPO is stable among the different energy bands in most exposures (see \autoref{fig:fre-energy} and \autoref{tab:all-fit-value}).

The fractional RMS of the new LFQPOs usually decreases with the increasing energy (\autoref{tab:all-fit-value} and \autoref{fig:rms-energy}), which is different from those of type-A, B, and C QPOs. Based on \textit{NICER} data, \cite{Alabarta_2022} reports that the fractional RMS of type-C QPOs increases up to 2-3 keV and then remains approximately constant above this energy. Similarly, as shown in \autoref{fig:rms-energy}, the fractional RMS of type-C QPOs generally exhibits higher values in the middle energy range. The fractional RMS of type-A,B QPOs both monotonically increase with energy at least up to 10 keV \citep{2020MNRAS.496.4366B,zhang_2023}, and maintain constant up to 100 keV for type-B QPO \citep{Liu_2022}.


\section{Conclusion AND DISCUSSION}
\label{sec:discussion}

We conducted a study on the X-ray timing properties of MAXI J1348$-$630 using the archival data from the \textit{Insight}-HXMT, and detected a new type LFQPO.
The characteristics of new LFQPO are different to previous classified  type-A,B,C QPOs, leading us to label it as an ``atypical" LFQPO. In this section, we summarize the distinct characteristics of the new LFQPO, and then discuss the feasibility to explain the new LFQPO with current  models for typical LFQPOs. Compared to typical LFQPOs, the new LFQPO shows the following characteristics:
\begin{enumerate}
\item Moderately stable frequency

The new LFQPO is detected during the rising and decaying phase of the main outburst and also during the mini outburst, the X-ray intensity spans more than one order of magnitude (\autoref{fig:lc-hid-rid}a). However, the frequency of the new LFQPO remains moderately stable at 0.08--0.15 Hz (\autoref{tab:all-fit-value}), and does not show an obvious variation with increasing/decreasing X-ray intensity. The new LFQPO does not follow the correlation between the QPO frequency and the BLN frequency (\autoref{fig:QPOfre-BLNfre}).

\item Exclusively detected in the hard state

From \autoref{fig:lc-hid-rid}a, we can obviously see that the new LFQPO only appears in the hard state, including the rising and decaying hard state of the main outburst and the hard state of the mini outburst. The new LFQPO is not detected in all exposures of hard state. The detection of the new LFQPO corresponds a very narrow range of X-ray hardness (\autoref{fig:lc-hid-rid}b).
Most detection of the new LFQPO accompany by the type-C QPO (\autoref{fig:fre-energy}), which is the most common LFQPO detected in the hard state.

\item Decreasing trend of fractional RMS dependence on energy

The fractional RMS of the new LFQPO seems decreases with energy in the 1--100 keV band (\autoref{fig:rms-energy}), which is different from the behavior of type-C and type-B QPOs.



\end{enumerate}

\subsection{``atypical" LFQPOs detection in other BHXRBs}
There are some other ``atypical" LFQPOs detected in BHXRBs, the mechanisms of which are still unclear. Some of them are thought to be related with the dip sources \citep[e.g.][]{Altamirano2012}, or high-inclination angle \citep{Huppenkothen2017}. However, the inclination angle of the jet of MAXI J1348$-$630 is about 29.3$\degree$ \citep{Carotenuto2022MNRAS}, which is thought to be aligned with the orbital plane. So the new LFQPO detected in MAXI J1348$-$630 should not share the same origin with those ones. There are LFQPOs at similar frequencies detected in BH high-mass X-ray binaries Cyg X-1 \citep[$\sim$0.09 Hz][]{Yan2021} and LMC X-1\citep[$\sim$0.08 Hz][]{Ebisawa1989}. The inclination angles of this two sources are also similar to MAXI J1348$-$630, 36$\degree$ for LMC X-1 and 27$\degree$ for Cyg X-1. However, their LFQPOs are detected in the soft state, in contrast the new LFQPOs of MAXI J1348$-$630 are exclusively detected in the hard state.

\subsection{Theoretical challenges in understanding the new LFQPO}

The existence of the new type LFQPO implies that, during BH accretion, different mechanisms related to the periodic processes works and exhibits quasi-periodic X-ray variability.
We will next discuss the challenges of generating such LFQPO within the framework of several models.

Many models have been proposed to explain the type-C QPO, which is the most common LFQPO detected in BHXRBs. The Lense-Thirring precession of the hot accretion flow is a popular model \citep[see reviews in][]{Ingram20}. The large span of central frequency
is one of the predominant characteristics of type-C QPO \citep{Belloni_2014}, which is clearly different to the new LFQPO (see \autoref{fig:QPOfre-BLNfre}).

For the truncated disc scenario, in which the hot accretion flow reside within the truncated thin disc, the BLN frequency depends on the truncated radius, i.e. the outer radius of the hot accretion flow \citep{Ingram11}, and therefore the QPO frequency is determined by the inner and outer radii of the hot accretion flow. As the truncated radius decreases with increasing X-ray luminosity, therefore, the frequencies of both type-C QPO and BLN increases, which naturally produce the positive correlation in \autoref{fig:QPOfre-BLNfre}. While for  the new LFQPO, the constant frequency seems unrelated to the truncation radius.
 
 
Jet precession was proposed to explain the soft phase lag of type-C QPO observed in MAXI J1820+070 \citep{Ma_2021} and the infrared QPO detected in GX 339-4 \citep{Kalamkar2016}. In this model, the QPO frequency is determined by the size of the jet or the radius of the rigidly coupled accretion flow, and the RMS depends on the jet velocity \citep{Liska2018,Ma_2021}. So an unchanged jet size and fine tuned jet velocities of different emission regions are required to explain the properties of the new LFQPO in this jet precession model.

Another model is proposed for the LFQPOs in XRBs is the accretion-ejection instability (AEI) in the accretion disc with a moderate magnetic field \citep{Tagger1999}. The QPO frequency predicted in this model also changes with the evolution of the inner radius of the accretion disc during the outburst \citep{Rodriguez2002,Varniere2002}. So the AEI model also can not explain the stable frequency of the new LFQPO.

One type of the discoseismic oscillations in the inner region of accretion disc: c-modes (or so-called corrugation waves) can naturally produce the oscillations corresponding to the frequency range of LFQPOs in XRBs with very low spinning BHs \citep{Kato2001,Silbergleit2001,Tsang2009MNRAS}. The BH spin of MAXI J1348$-$630 is around 0.8 \citep{Kumar2022,Song2023}, which is not nearly low enough to produce $\sim$0.1 Hz oscillation \citep{Silbergleit2001}.

An oscillating corona model has been proposed to explain the LFQPOs in XRBs \citep{Cabanac10}. The QPO frequency will be around 0.1 Hz if the size of the corona is about hundreds of $R_\mathrm{g}$. However, the fractional RMS predicted in this model increases with energy, which is opposite to the behavior observed in the new LFQPO detected in MAXI J1348$-$630.

To conclude, non of the above mentioned models can explain all the observed properties of the new LFQPO. Considering its moderately stable frequency, the oscillation should originate from a standing component or at fixed radius of the accretion flow. As strong reflection can produce the decreasing trend of fractional RMS dependence on increasing energy\citep[e.g.][]{Gao_2023}, the QPO should be related with the cold disc rather than the corona or the hot accretion flow. However, the reflection components of MAXI J1348$-$630 during hard state are relatively weak \citep{Jia_2022}. To explore the origin of the new LFQPO,  further investigation is needed.


\section*{acknowledgements}
\label{sec:ack}
This work made use of the publicly available data and software from the \textit{Insight}-HXMT mission. The \textit{Insight}-HXMT project is funded by the China National Space Administration (CNSA) and the Chinese Academy of Sciences (CAS). 
X.L.W. \& R.Y.M. were supported by the National Key R\&D Program of China (Grant No. 2023YFA1607902);
F.G.X. \& R.Y.M. were supported by the National SKA Program of China (no. 2020SKA0110102);
J.F.W. was supported by the National Key R\&D Program of China (Grant no. 2023YFA1607904);
Z.Y. was supported by the Youth Innovation Promotion Association of Chinese Academy of Sciences (YIPA-CAS);
This works was supported in part by the Natural Science Foundation of China (NSFC, grants U1838203,U1938114,12373049, 12361131579, 12373017, 12192220, 12192223, 12033004, 12221003, U2038108 and 12133008).

\section*{Data Availability}
The data underlying this article are available in the \textit{Insight}-HXMT public archive.


\appendix
\section{The fitting results of the new LFQPO in other exposures}
\label{appendix:all-fit-value}

\begin{table*}
 \centering
  \caption{The fitting parameters of the new LFQPO from the other exposures. Some errors are absent because we fixed the quality factor of the QPO when fitting energy bands with lower signal-to-noise ratios.}
    \begin{tabular}{rrllll}
    \toprule
 \multicolumn{1}{c}{Exposure ID} & \multicolumn{1}{c}{MJD} & \multicolumn{1}{l}{Parameter} & \multicolumn{1}{c}{HXMT LE} & \multicolumn{1}{c}{HXMT ME} & \multicolumn{1}{c}{HXMT HE} \\
   \midrule
    \multicolumn{1}{c}{P021400200403} & \multicolumn{1}{c}{58513.60} & Frequency (Hz) & $0.102^{+0.023}_{-0.028}$ & $0.095^{+0.003}_{-0.003}$ & $0.098^{+0.007}_{-0.007}$ \\
          &       & FWHM (Hz) & $0.003^{+0.006}_{-0.002}$ & $0.01^{+0.0}_{-0.0}$ & $0.01^{+0.0}_{-0.0}$ \\
          &       & RMS (\%) & $3.1^{+1.1}_{-1.1}$ & $1.8^{+0.35}_{-0.35}$ & $0.92^{+0.27}_{-0.32}$ \\
    \midrule
    \multicolumn{1}{c}{P021400200410} & \multicolumn{1}{c}{58514.54} & Frequency (Hz) & $0.137^{+0.004}_{-0.003}$ & $0.149^{+0.005}_{-0.006}$ & $0.135^{+0.003}_{-0.004}$ \\
          &       & FWHM (Hz) & $0.006^{+0.009}_{-0.005}$ & $0.017^{+0.018}_{-0.013}$ & $0.015^{+0.011}_{-0.009}$ \\
          &       & RMS (\%) & $3.3^{+1.0}_{-1.3}$ & $1.7^{+0.48}_{-0.57}$ & $1.29^{+0.27}_{-0.32}$ \\
    \midrule
    \multicolumn{1}{c}{P021400200416} & \multicolumn{1}{c}{58515.34} & Frequency (Hz) & $0.119^{+0.007}_{-0.006}$ & $0.095^{+0.004}_{-0.004}$ & $0.111^{+0.006}_{-0.007}$ \\
          &       & FWHM (Hz) & $0.022^{+0.029}_{-0.016}$ & $0.006^{+0.007}_{-0.005}$ & $0.052^{+0.024}_{-0.016}$ \\
          &       & RMS (\%) & $2.9^{+0.9}_{-1.5}$ & $1.49^{+0.47}_{-0.53}$ & $2.1^{+0.28}_{-0.29}$ \\
    \midrule
    \multicolumn{1}{c}{P021400207302} & \multicolumn{1}{c}{58610.87} & Frequency (Hz) & $0.091^{+0.007}_{-0.006}$ & $0.141^{+0.004}_{-0.002}$ & $0.104^{+0.003}_{-0.003}$ \\
          &       & FWHM (Hz) & $0.039^{+0.039}_{-0.019}$ & $0.01^{+0.011}_{-0.006}$ & $0.011^{+0.013}_{-0.007}$ \\
          &       & RMS (\%) & $3.1^{+0.7}_{-1.2}$ & $1.18^{+0.21}_{-0.26}$ & $0.79^{+0.18}_{-0.21}$ \\
    \midrule
    \multicolumn{1}{c}{P020101200701} & \multicolumn{1}{c}{58644.23} & Frequency (Hz) & $0.102^{+0.004}_{-0.004}$ & $0.13^{+0.008}_{-0.008}$ & $0.135^{+0.006}_{-0.005}$ \\
          &       & FWHM (Hz) & $0.008^{+0.01}_{-0.005}$ & $0.043^{+0.0}_{-0.0}$ & $0.027^{+0.0}_{-0.0}$ \\
          &       & RMS (\%) & $3.5^{+0.9}_{-1.0}$ & $2.63^{+0.42}_{-0.37}$ & $1.69^{+0.27}_{-0.24}$ \\
    \midrule
    \multicolumn{1}{c}{P021400208001} & \multicolumn{1}{c}{58645.02} & Frequency (Hz) &       & $0.123^{+0.003}_{-0.005}$ & $0.136^{+0.009}_{-0.013}$ \\
          &       & FWHM (Hz) &       & $0.025^{+0.0}_{-0.0}$ & $0.01^{+0.024}_{-0.008}$ \\
          &       & RMS (\%) &       & $2.75^{+0.42}_{-0.42}$ & $1.35^{+0.47}_{-0.56}$ \\
    \midrule
    \multicolumn{1}{c}{P021400208201} & \multicolumn{1}{c}{58648.47} & Frequency (Hz) & $0.132^{+0.004}_{-0.004}$ & $0.127^{+0.005}_{-0.005}$ & $0.125^{+0.004}_{-0.004}$ \\
          &       & FWHM (Hz) & $0.021^{+0.012}_{-0.01}$ & $0.013^{+0.0}_{-0.0}$ & $0.012^{+0.0}_{-0.0}$ \\
          &       & RMS (\%) & $4.05^{+0.63}_{-0.93}$ & $1.59^{+0.48}_{-0.49}$ & $1.23^{+0.29}_{-0.37}$ \\
    \midrule
    \multicolumn{1}{c}{P021400208301} & \multicolumn{1}{c}{58649.07} & Frequency (Hz) & $0.137^{+0.003}_{-0.003}$ & $0.146^{+0.008}_{-0.009}$ & $0.139^{+0.008}_{-0.008}$ \\
          &       & FWHM (Hz) & $0.005^{+0.006}_{-0.004}$ & $0.038^{+0.026}_{-0.018}$ & $0.049^{+0.027}_{-0.021}$ \\
          &       & RMS (\%) & $4.3^{+1.4}_{-1.8}$ & $2.97^{+0.74}_{-0.54}$ & $2.47^{+0.43}_{-0.58}$ \\
    \bottomrule
    \end{tabular}%
  \label{tab:all-fit-value}%
\end{table*}%

\begin{table*}
 \centering
  \caption*{}
    \begin{tabular}{rrllll}
    \midrule
    \multicolumn{1}{c}{P021400208401} & \multicolumn{1}{c}{58650.66} & Frequency (Hz) & $0.107^{+0.006}_{-0.005}$ & $0.106^{+0.003}_{-0.002}$ & $0.1^{+0.004}_{-0.004}$ \\
          &       & FWHM (Hz) & $0.021^{+0.021}_{-0.014}$ & $0.006^{+0.008}_{-0.004}$ & $0.014^{+0.011}_{-0.008}$ \\
          &       & RMS (\%) & $3.8^{+1.0}_{-1.3}$ & $1.91^{+0.48}_{-0.46}$ & $2.27^{+0.53}_{-0.45}$ \\
    \midrule
    \multicolumn{1}{c}{P021400208402} & \multicolumn{1}{c}{58650.81} & Frequency (Hz) & $0.149^{+0.004}_{-0.005}$ & $0.149^{+0.006}_{-0.008}$ & $0.146^{+0.006}_{-0.011}$ \\
          &       & FWHM (Hz) & $0.015^{+0.0}_{-0.0}$ & $0.008^{+0.013}_{-0.006}$ & $0.019^{+0.021}_{-0.013}$ \\
          &       & RMS (\%) & $4.21^{+0.86}_{-0.81}$ & $1.4^{+0.38}_{-0.48}$ & $1.2^{+0.32}_{-0.44}$ \\
    \midrule
    \multicolumn{1}{c}{P021400208602} & \multicolumn{1}{c}{58652.80} & Frequency (Hz) & $0.129^{+0.002}_{-0.002}$ & $0.12^{+0.015}_{-0.02}$ & $0.098^{+0.002}_{-0.003}$ \\
          &       & FWHM (Hz) & $0.012^{+0.007}_{-0.005}$ & $0.035^{+0.079}_{-0.028}$ & $0.011^{+0.011}_{-0.008}$ \\
          &       & RMS (\%) & $4.63^{+0.76}_{-0.69}$ & $1.26^{+0.52}_{-0.97}$ & $1.46^{+0.39}_{-0.43}$ \\
    \midrule
    \multicolumn{1}{c}{P020101200803} & \multicolumn{1}{c}{58654.11} & Frequency (Hz) & $0.123^{+0.013}_{-0.014}$ & $0.115^{+0.011}_{-0.01}$ & $0.127^{+0.006}_{-0.006}$ \\
          &       & FWHM (Hz) & $0.05^{+0.038}_{-0.029}$ & $0.029^{+0.0}_{-0.0}$ & $0.038^{+0.0}_{-0.0}$ \\
          &       & RMS (\%) & $5.4^{+1.3}_{-1.7}$ & $2.14^{+0.53}_{-0.59}$ & $2.11^{+0.32}_{-0.32}$ \\
    \midrule
    \multicolumn{1}{c}{P020101200901} & \multicolumn{1}{c}{58654.63} & Frequency (Hz) & $0.096^{+0.007}_{-0.006}$ & $0.083^{+0.003}_{-0.003}$ & $0.103^{+0.003}_{-0.003}$ \\
          &       & FWHM (Hz) & $0.03^{+0.0}_{-0.0}$ & $0.006^{+0.007}_{-0.004}$ & $0.031^{+0.0}_{-0.0}$ \\
          &       & RMS (\%) & $2.71^{+0.91}_{-0.97}$ & $1.78^{+0.46}_{-0.54}$ & $1.72^{+0.3}_{-0.32}$ \\
    \midrule
    \multicolumn{1}{c}{P020101200902} & \multicolumn{1}{c}{58654.79} & Frequency (Hz) & $0.096^{+0.01}_{-0.012}$ & $0.114^{+0.011}_{-0.012}$ & $0.097^{+0.004}_{-0.004}$ \\
          &       & FWHM (Hz) & $0.055^{+0.042}_{-0.024}$ & $0.034^{+0.0}_{-0.0}$ & $0.01^{+0.0}_{-0.0}$ \\
          &       & RMS (\%) & $5.9^{+1.6}_{-2.2}$ & $2.12^{+0.4}_{-0.49}$ & $1.05^{+0.32}_{-0.32}$ \\
    \midrule
    \multicolumn{1}{c}{P020101201101} & \multicolumn{1}{c}{58657.02} & Frequency (Hz) & $0.109^{+0.004}_{-0.003}$ & $0.12^{+0.005}_{-0.005}$ & $0.112^{+0.005}_{-0.004}$ \\
          &       & FWHM (Hz) & $0.009^{+0.023}_{-0.007}$ & $0.024^{+0.0}_{-0.0}$ & $0.012^{+0.014}_{-0.008}$ \\
          &       & RMS (\%) & $3.8^{+1.1}_{-1.5}$ & $2.78^{+0.43}_{-0.42}$ & $1.52^{+0.36}_{-0.46}$ \\
    \midrule
    \multicolumn{1}{c}{P020101201103} & \multicolumn{1}{c}{58657.30} & Frequency (Hz) & $0.092^{+0.005}_{-0.004}$ & $0.107^{+0.004}_{-0.005}$ & $0.107^{+0.005}_{-0.004}$ \\
          &       & FWHM (Hz) & $0.032^{+0.013}_{-0.01}$ & $0.022^{+0.021}_{-0.012}$ & $0.018^{+0.016}_{-0.01}$ \\
          &       & RMS (\%) & $6.1^{+1.1}_{-0.9}$ & $2.35^{+0.58}_{-0.81}$ & $1.79^{+0.41}_{-0.46}$ \\
    \midrule
    \multicolumn{1}{c}{P020101201303} & \multicolumn{1}{c}{58661.54} & Frequency (Hz) & $0.091^{+0.002}_{-0.002}$ & $0.096^{+0.002}_{-0.003}$ & $0.095^{+0.009}_{-0.022}$ \\
          &       & FWHM (Hz) & $0.004^{+0.005}_{-0.003}$ & $0.003^{+0.005}_{-0.003}$ & $0.003^{+0.007}_{-0.003}$ \\
          &       & RMS (\%) & $3.18^{+0.89}_{-0.86}$ & $2.13^{+0.48}_{-0.46}$ & $1.7^{+0.5}_{-0.53}$ \\
    \bottomrule
    \end{tabular}%
\end{table*}%

\begin{table*}
 \centering
  \caption*{}
    \begin{tabular}{rrllll}
    \midrule
    \multicolumn{1}{c}{P020101201402} & \multicolumn{1}{c}{58663.33} & Frequency (Hz) & $0.093^{+0.015}_{-0.015}$ & $0.094^{+0.006}_{-0.008}$ & $0.094^{+0.018}_{-0.019}$ \\
          &       & FWHM (Hz) & $0.025^{+0.0}_{-0.0}$ & $0.047^{+0.0}_{-0.0}$ & $0.006^{+0.01}_{-0.004}$ \\
          &       & RMS (\%) & $3.5^{+1.0}_{-1.2}$ & $3.17^{+0.63}_{-0.62}$ & $1.26^{+0.45}_{-0.73}$ \\
    \midrule
    \multicolumn{1}{c}{P020101201802} & \multicolumn{1}{c}{58672.43} & Frequency (Hz) & $0.082^{+0.005}_{-0.004}$ & $0.077^{+0.004}_{-0.004}$ & $0.082^{+0.001}_{-0.001}$ \\
          &       & FWHM (Hz) & $0.008^{+0.0}_{-0.0}$ & $0.02^{+0.015}_{-0.011}$ & $0.005^{+0.008}_{-0.004}$ \\
          &       & RMS (\%) & $2.8^{+0.9}_{-1.1}$ & $2.19^{+0.5}_{-0.6}$ & $1.56^{+0.34}_{-0.41}$ \\
    \bottomrule
    \bottomrule
    \end{tabular}%
\end{table*}%



\bibliographystyle{aasjournal}
\bibliography{ms} 





\label{lastpage}
\end{document}